\documentclass[12pt]{iopart}
\usepackage{graphicx}
\begin{document}
\title{Analytical study on the criticality of the Stochastic Optimal Velocity model}
\author{Masahiro Kanai$^1$, Katsuhiro Nishinari$^2$, Tetsuji Tokihiro$^1$}%
\address{$^1$ Graduate School of Mathematical Sciences, 
University of Tokyo, 3-8-1 Komaba, Tokyo 153-8914, Japan}
\address{$^2$ Department of Aeronautics and Astronautics, Faculty of Engineering, University of Tokyo, 7-3-1 Hongo, Tokyo 113-8656, Japan}
\ead{kanai@ms.u-tokyo.ac.jp}
\begin{abstract}
In recent works, we have proposed a stochastic cellular automaton
 model of traffic flow connecting two exactly solvable
 stochastic processes, i.e., the Asymmetric Simple Exclusion Process
 and the Zero Range Process, with an additional parameter.
It is also regarded as an extended version
 of the Optimal Velocity model,
 and moreover it shows particularly notable properties.
In this paper, we report that when taking Optimal Velocity function
 to be a step function, all of the flux-density graph
 (i.e. the fundamental diagram) can be estimated.
We first find that the fundamental diagram consists of
 two line segments resembling an {\it inversed-$\lambda$} form,
 and next identify their end-points from a microscopic behaviour
 of vehicles.
It is notable that by using a microscopic parameter
 which indicates a driver's sensitivity to the traffic situation,
 we give an explicit formula for the critical point
 at which a traffic jam phase arises.
We also compare these analytical results
 with those of the Optimal Velocity model,
 and point out the crucial differences between them.
\end{abstract}
\submitto{\JPA}
\maketitle
\section{Introduction}
Traffic dynamics has naturally attracted much attention
 from engineers since the volume of vehicular traffic outstripped
 road capacity \cite{helbing,css,nagatani}.
Especially for the last decades, it has attracted a lot of interest,
 in addition, from physicists and mathematicians
 as a typical example of non-equilibrium statistical mechanics
 of self-driven many particle systems \cite{helbing}.
Since self-driven particles do not obey Newton's laws of motion,
 their collective phenomena are far from predictable
 and are strongly dependent on their density.
Recently, a number of different approaches to the subject
 have been made from various viewpoints
 such as microscopic and macroscopic,
 continuous and discrete, deterministic and stochastic.

The Burgers equation $\rho_t=2\rho\rho_x+\rho_{xx}$,
 is known as an elementary model of traffic flow.
It formulates the evolution of density distribution of vehicles,
 i.e., it is a macroscopic, continuous model.
Recent studies show that the Burgers equation is directly connected
 with other basic models of traffic flow \cite{Matsukidaira}.
The Burgers equation is, at first, transformed
 into a cellular automaton (CA) model (a microscopic discrete one)
 through {\it ultra-discretization} \cite{Tokihiro}.
Note that
 CA models are microscopic ones because they explicitly define
 each particle's motion.
It provides a method to reveal another profile of models,
 and enables us to return from macroscopic to microscopic.
From the Burgers equation, we thereby obtain the so-called
 {\it Burgers cellular automata} (BCA),
 which is an extension of the Rule-184 CA \cite{Matsukidaira}.
This CA has a very simple update rule, i.e.,
 if the adjacent site is not occupied,
 the vehicle move ahead with a given probability,
 but otherwise it does not.
Each site contains one vehicle at most,
 and the rule is generally called
 {\it the hard-core exclusion rule}.
In the case of BCA, each site can contain more than one vehicle.

Moreover, BCA can be transformed into another basic model, i.e.,
 the {\it Optimal Velocity (OV) model},
 through the {\it discrete Euler-Lagrange (E-L) transformation}.
The discrete E-L transformation is made
 on fully discrete variables,
 and then field variables change to particle variables
 \cite{Matsukidaira}.
The OV model is a continuous, microscopic model, and is expressed as
\begin{equation}
\ddot{x}_i=a\Bigl[V(x_{i+1}-x_i)-\dot{x}_i\Bigr],\label{OV}
\end{equation}
 where $x_i=x_i(t)$ denotes the position of the $i$-th vehicle
 at time $t$ and the function $V$ is called
 the {\it Optimal Velocity (OV) function} \cite{Bando,Bando2}.
The OV function gives the optimal velocity of a vehicle
 in terms of the headway $x_{i+1}-x_i$,
 where the $i$-th vehicle follows the $(i+1)$-th in the same lane,
 and then the OV function, in general, is monotonically increasing.
In particular, the OV model obtained from BCA has OV function
 which is a step function.
This suggests that the OV model with a step function
 is essential as well as elementary.

Cellular automaton models are efficient and flexible
 compared to those described by differential equations,
 and they have been used to model complex traffic systems
 such as ramps and crossings \cite{css}.
The Nagel-Schreckenberg (N-S) model, a well-known CA model,
 successfully reproduces typical properties of real traffic \cite{NS}.
What makes it sophisticated is a {\it random braking rule},
 which is a plausible mechanism to simulate the motion of vehicles
 in a single lane.
Nevertheless, the N-S model does not succeed in reproducing
 the so-called {\it metastable state}, i.e., an unstable state
 which breaks down to the lower-flux stable state
 under some perturbations.

Extensive study of traffic flow has revealed
 that the metastable property, appearing in the medium density region,
 is universal in real traffic flows, and accordingly
 that property is essential in modelling \cite{Kerner,NFS}.
Moreover, it is quite distinct
 among non-equilibrium statistical systems \cite{helbing}.
In other words, this metastable property plays a critical role
 in characterizing traffic flow from the viewpoint of dynamics,
 and it is hence required for traffic models to exhibit this property.
Thus far, one needs the {\it slow-start rule} to reproduce
 a metastable state in existing CA models \cite{NFS,TT,SS,Ni01}.
It introduces a delay for vehicles to respond
 to the changing traffic situation,
 i.e., if a vehicle stops due to the hard-core exclusion rule,
 the slow-start rule forces it to stop again at the next time step.
In contrast, due to the second-order derivative,
 the OV model naturally includes a similar mechanism
 to the slow-start rule.
The intrinsic parameter $a$ in the OV model (\ref{OV})
 corresponds to the driver's sensitivity to a traffic situation
 (e.g. the distances relatively to the vehicles ahead),
 and plays an important role in the stability of a traffic flow.
Note that the reciprocal of $a$ represents a driver's response time,
 connecting the OV model and the Newell model \cite{Newell}.

In the next section, we introduce a stochastic CA model
 following \cite{KNT}, and then we show that
 the model inherits the sophisticated features from the OV model.
\section{Stochastic optimal velocity model}
Most stochastic models incorporate noise, taking into account
 uncertain effects such as different driver characteristics,
 driver error false operation, and external influences.
In general, randomness disturbs metastable states
 and thus stochastic models do not show any metastable state
 in the fundamental diagram.
In contrast, as will be seen in the latter part of this paper,
 the probability of the SOV model plays an essential role
 in producing metastable states, and the metastable states emerge
 by extracting a deterministic mechanism.
Note that a similar trick was considered in \cite{appert}.
\subsection{General scheme}
First of all, we explain the general framework
 of our stochastic CA model for one-lane traffic.
The roadway, being divided into cells, is regarded
 as a one-dimensional array of $L$ sites,
 and each site contains one vehicle at most.
Let ${\sf M}^t_i$ be a stochastic variable
 which denotes the number of sites
 through which the $i$-th vehicle moves at time $t$,
 and $w^t_i(m)$ be the probability
 that ${\sf M}^t_i=m~(m=0,1,2,\ldots)$.
Then, we assume a principle of motion
 that the probability $w^{t+1}_i(m)$ depends on
 the probability distribution $w^t_i(0),w^t_i(1),\ldots$,
 and the positions of vehicles $x^t_1,x^t_2,\ldots,x^t_N$
 at the previous time.
The updating procedure is as follows:
\begin{itemize}
\item Calculate the next intention $w^{t+1}_i~(i=1,2,\ldots,N)$
 from the present, intention $w^t_i(0),w^t_i(1),\ldots$
 and positions $x^t_1,x^t_2,\ldots,x^t_N$;
\begin{equation}
w^{t+1}_i(m)=f(w^t_i(0),w^t_i(1),\ldots;x^t_1,\ldots,x^t_N;m)\label{gen}
\end{equation}
\item Determine the number of sites ${\sf M}^{t+1}_i$
 through which a vehicle moves (i.e. the velocity) probabilistically
 according to the intention $w^{t+1}_i$.
\item The new position of each vehicle is
\begin{equation}
x^{t+1}_i=x^t_i+\min(\Delta x^t_i,\,{\sf M}^{t+1}_i)\quad(\forall i),\label{genx}
\end{equation}
where $\Delta x^t_i=x^t_{i+1}-x^t_i-1$ denotes the headway.
(Headway is defined to be the clear space in front of the vehicle,
 and thus in a CA model we need to subtract 1 to take account of
 the site occupied by the vehicle itself.)
\end{itemize}
The hard-core exclusion rule is incorporated
 through the second term of the right hand side of (\ref{genx}).

We call the probability distribution $w^t_i$ {\it the intention}
 because it is an intrinsic variable of the vehicle
 and drives themselves.
It brings uncertainty of operation into the traffic model
 and has no physical counterpart.
\subsection{The SOV model}
In what follows, we assume $w^t_i(m)\equiv0$ for $m\geq 2$.
It is notable that $\sum^\infty_{m=0}w^t_i(m)=1$ by definition
 and the expectation value
 $\langle M^t_i\rangle=\sum^\infty_{m=0}mw^t_i(m)$,
 and hence, setting $v^t_i=w^t_i(1)$,
 we have $w^t_i(0)=1-v^t_i$ and $\langle M^t_i\rangle=v^t_i$.
{From} (\ref{gen}) we have
\begin{equation}
\left\{
\begin{array}{l}
w^{t+1}_i(1)=v^{t+1}_i=f(v^t_i;x^t_1,x^t_2,\ldots;1)\\
w^{t+1}_i(0)=1-v^{t+1}_i,
\end{array}\label{gen2}
\right.
\end{equation}
 and we therefore express the intention by $v^t_i$ in stead of $w^t_i$.
As long as vehicles move separately (i.e. $\Delta x^t_i\gg0$),
 the positions are updated according to the simple form
\begin{equation}
x^{t+1}_i=\left\{
\begin{array}{ll}
x^t_i+1\quad & \mbox{with probability}~~v^{t+1}_i\\
x^t_i & \mbox{with probability}~~1-v^{t+1}_i,
\end{array}
\right.
\end{equation}
 and consequently we have
\begin{equation}
\langle x^{t+1}_i\rangle=\langle x^t_i\rangle+v^{t+1}_i\label{SOVx}
\end{equation}
 in the sense of expectation value.
This equation expresses the fact that the intention $v^{t+1}_i$ can be
 regarded as the average velocity at time $t$.

Let us take an evolution equation
\begin{equation}
v^{t+1}_i=(1-a)v^t_i+aV(\Delta x^t_i),\label{SOV}
\end{equation}
 in (\ref{gen}), where $a~(0\leq a\leq 1)$ is a parameter
 and the function $V$ takes the value in $[0,1]$
 so that $v^t_i$ should be within $[0,1]$.
Equation (\ref{SOV}) consists of two terms, i.e.,
 a term turning over the intention $v^t_i$ into the next,
 and an effect of the situation (the headway $\Delta x^t_i$).
The intrinsic parameter $a$ indicates the sensitivity of vehicles
 to the traffic situation, and the larger $a$ is,
 the less time a vehicle takes to change its intention.

A discrete version of the OV model is expressed as
\begin{eqnarray}
x_i(t+\Delta t)-x_i(t)&=&v_i(t)\Delta t\label{dOVx},\\
v_i(t+\Delta t)-v_i(t)&=&a\Bigl[V(\Delta x_i(t))-v_i(t)\Bigr]\label{dOVv}\Delta t,
\end{eqnarray}
 where $\Delta x_i(t)=x_{i+1}(t)-x_i(t)$,
 and $\Delta t$ is a time interval.
Due to the formal correspondence
 between (\ref{SOV}) and (\ref{dOVv}),
 we call a stochastic CA model defined by (\ref{SOV})
 the {\it Stochastic Optimal Velocity (SOV) model}, hereafter.

As we noted in the preceding work \cite{KNT},
 the SOV model reduces to two exactly solvable stochastic models,
 the Asymmetric Simple Exclusion Process (ASEP)
 \cite{Rajewsky,Schutz}
 and the Zero Range Process (ZRP) \cite{Spitzer,Evans1},
 when the parameter $a~(0\leq a\leq1)$ takes
 the values of $0$ and $1$ respectively.
The SOV model has a fundamental diagram similar to that of ZRP
 as $a$ approaches 1 (the ZRP limit).
However, when $a$ takes a small value (the ASEP limit), the SOV model
 shows quite different properties from those of ASEP.
Since we already discussed this point in detail \cite{KNT},
 we do not devote any space to the limiting case $a\rightarrow0$.
It should be noted here that a model including ZRP and ASEP
 was proposed in \cite{Klauck}.
However, this model does not satisfy the essential standard
 that a traffic model should reproduce the metastable state
 observed universally in empirical traffic data.
\section{The SOV model with a step OV function}
In this section, we take a step function
\begin{equation}
V(x)=\left\{
\begin{array}{ll}
0&\qquad (0\leq x<d)\\
1&\qquad (x\geq d)
\end{array}\label{step}
\right.
\end{equation}
 as the OV function.
Then, we impose a periodic boundary condition
 and adopt the parallel updating as usual.
\subsection{Comparison with the OV model}
\begin{figure}[t]
\begin{center}
\includegraphics[scale=0.7]{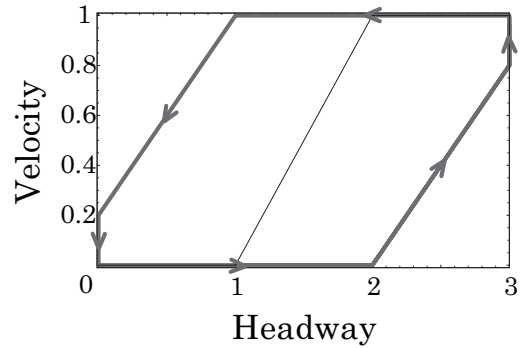}
\caption{
The headway-velocity diagram (phase space) of the SOV model ($a=0.8$)
 with a step function ($d=2$).
We observe a hysteresis loop (thick gray line)
 around the discontinuous point of the step OV function
 (thin black line),
 where the phase of a vehicle goes round in the direction of arrows.
In comparison with Fig. 3 of \cite{Sugiyama},
 two corners of the parallelogram are folded
 by an effect of the hard-core exclusion rule.
}
\label{hv}
\end{center}
\end{figure}
\begin{figure}[t]
\begin{center}
\includegraphics[scale=0.7]{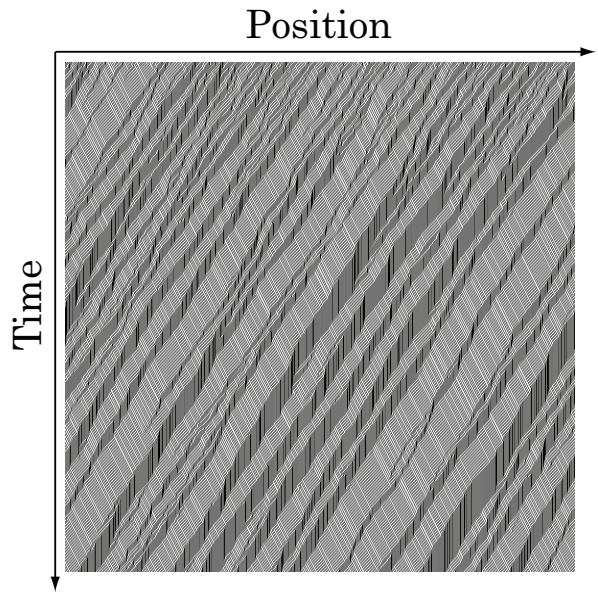}
\caption{
The spatio-temporal pattern of the SOV model ($a=0.8$)
 with a step function ($d=2$) simulated
 with the number of site $L=1000$
 and the number of vehicles $M=400$,
 where periodic boundary condition is imposed on the roadway.
There appear a lot of stable clusters (small jams)
 propagating backward.
}
\label{stp}
\end{center}
\end{figure}
In preceding works \cite{Sugiyama,Nakanishi},
 the original OV model with the step OV function (\ref{step})
 has been studied in detail.
In an ideal case, they gave an exact solution for cluster formation
 (i.e. the traveling cluster solution).
They use figures to illustrate the behaviour of vehicles
 forming a cluster, and find a closed loop or {\it hysteresis loop}
 in the velocity-headway diagram (phase space).
Moreover, from the microscopic viewpoint, they evaluate
 the velocity of a jam transmitting backward,
 and two specific headways;
 the distance $\Delta x_J$ with which vehicles stop in a jam,
 and $\Delta x_F$ with which vehicles move in a free-flow region.
(See \S\ref{criticalpoint}.)

Figure~\ref{hv} shows the phase space of our model.
Note that the vertical axis of Figure~\ref{hv}
 means the expectation value of velocity.
We find that there appears a hysteresis loop
 corresponding to Figure 3 in \cite{Sugiyama}.
Since our model incorporates the hard-core exclusion rule
 as well as a CA model with maximum allowed velocity 1,
 it takes a different shape from the parallelogram of Figure 3
 in \cite{Sugiyama}.
Note that, in contrast with the original OV model,
 our model contains the element of randomness and hence
 the vehicles do not always move in accordance with the hysteresis loop.
In Figure~\ref{stp}, we show the spatio-temporal pattern of our model,
 which should be compared with Figure~4 in \cite{Sugiyama}.
Due to the randomness, the clusters do not hold their sizes constant
 and the number of them fluctuates with time.

These two figures suggest that the SOV model inherits,
 from the original OV model, the mechanism of vehicles clustering
 and then separating.
However, from the viewpoint of many-particle systems,
 the SOV model shows apparently different collective phenomena
 from that of the OV model.
In the following part of this section,
 we estimate the fundamental diagram of the SOV model
 with a step function
 as well as comparing the estimated specific headways and velocity
 of our model with those given in \cite{Sugiyama}.
\subsection{Fundamental diagram}
We denote the density of vehicles to sites by $\rho=N/L$
 ($L$ is the number of sites, and $N$ the number of vehicles),
 which is a macroscopic variable, and a conserved quantity of motion
 under the periodic boundary condition, no entrances or exits.
Another macroscopic variable flux, $Q=\rho v$, is defined using
 the average velocity in a steady state;
\begin{equation}
v:=\frac1N\sum^N_{i=1}(x^T_i-x^{T-1}_i),
\end{equation}
where time $T$ should be taken large enough for the system
 to reach a steady state.

A {\it fundamental diagram}, a plot of the flux versus the density,
 illustrates how traffic conditions depend on density.
It represents the characteristics of a traffic model,
 and hence traffic models are required to reproduce
 a fundamental diagram observed in real traffic flow.
As far as the above-mentioned exactly solvable models
 (ASEP and ZRP) are concerned, we can make an exact calculation
 of the fundamental diagram \cite{SSNI,Evans2}.
We only exhibit an explicit formula of the fundamental diagram of ZRP
 (i.e. $a=1$) especially when a step function is adopted as the OV function:
\begin{equation}
Q^{\sf ZRP}(\rho)=
\left\{
\begin{array}{ll}
\min(\rho,1-d\rho)&\qquad(0\leq\rho\leq\frac1d)\\
0&\qquad(\frac1d< \rho\leq 1)
\end{array}
\right.\label{FluxOfZRP}
\end{equation}
where $d$ is the discontinuous point of
 the step OV function (\ref{step}).

Since the flux of R184-CA
 (ASEP with the hopping probability $p=1$)
 is obtained in the simple form of $Q(\rho)=\min(\rho,1-\rho)$,
 considering that each vehicle occupies $d$ sites,
 we directly obtain the formula (\ref{FluxOfZRP}).

\subsection{Metastable state}
\begin{figure}[t]
\begin{center}
\includegraphics[scale=0.5]{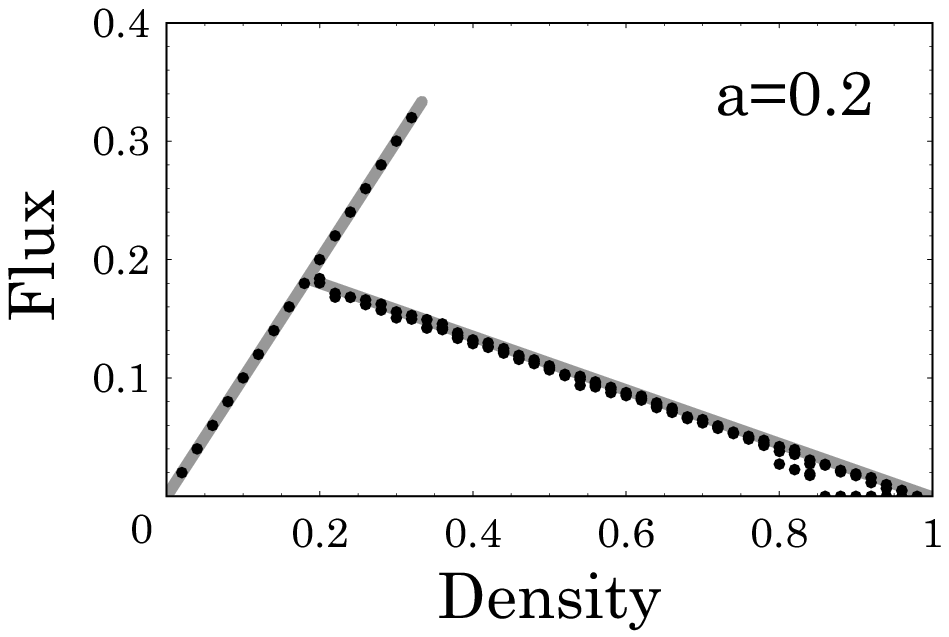}
\hspace{5pt}
\includegraphics[scale=0.5]{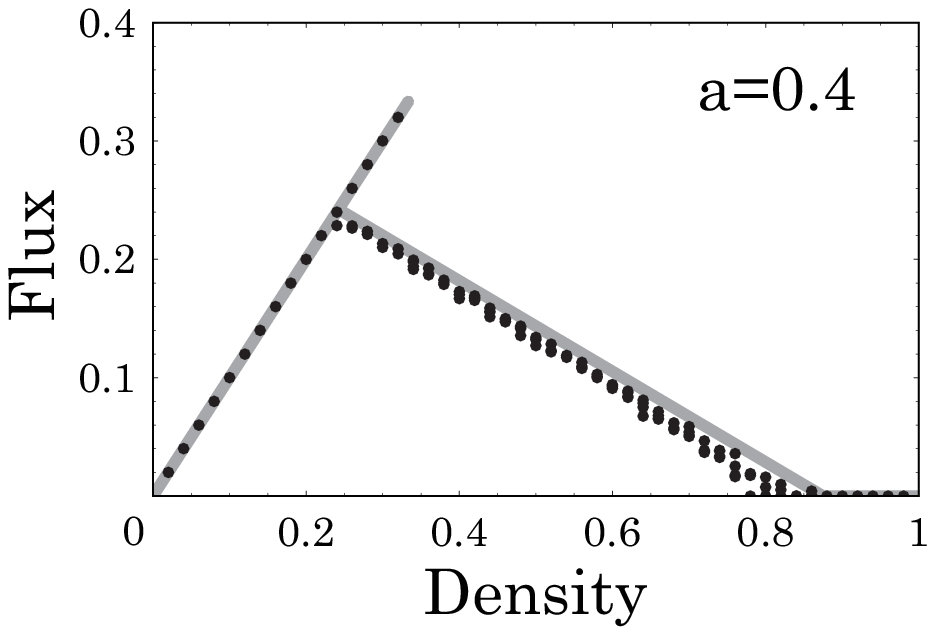}\\[5pt]
\includegraphics[scale=0.5]{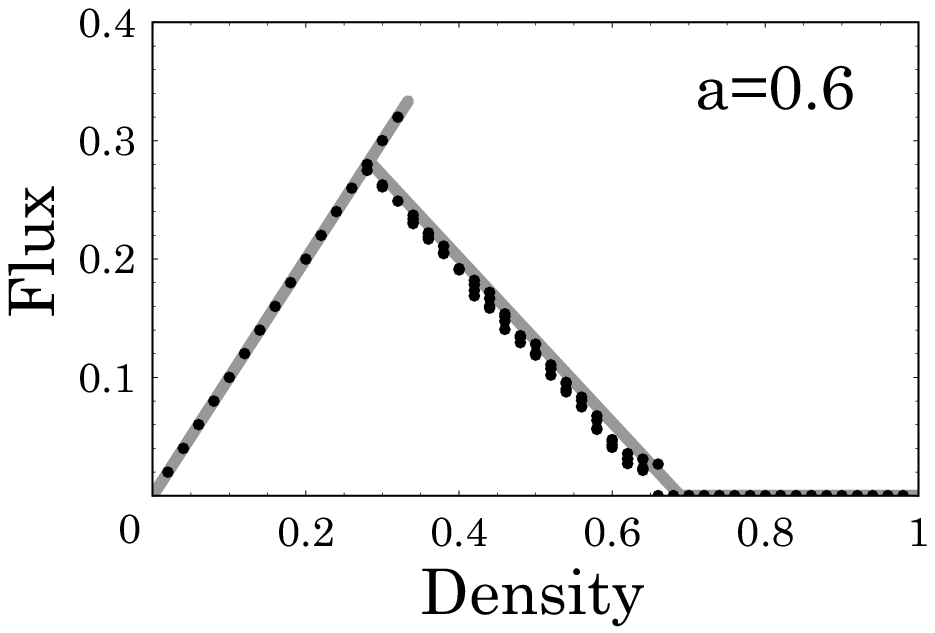}
\hspace{5pt}
\includegraphics[scale=0.5]{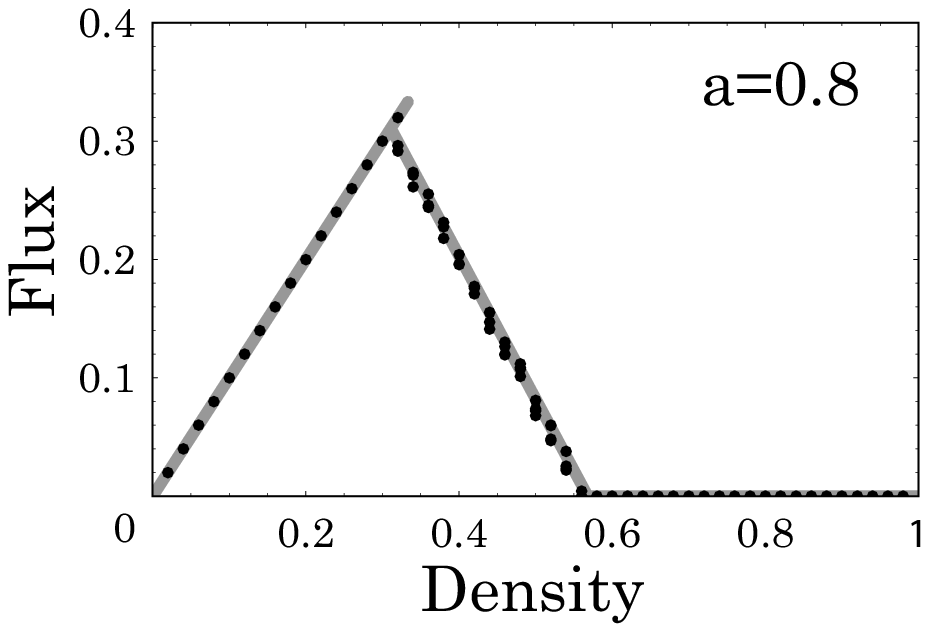}
\caption{
The fundamental diagram of the SOV model
 with the step OV function (\ref{step}) plotted
 at each value of sensitivity parameter $a$,
 where the discontinuous point is $d=2$,
 the initial value of the intention is $v^0_i=1$,
 and the system size is $L=1000$.
Theoretical curve (gray) has a complete agreement
 with simulated data (dots) in these cases.
}\label{fd}
\end{center}
\end{figure}
Figure~\ref{fd} shows the fundamental diagram of the SOV model
 with the step OV function (\ref{step}).
We find that the diagram consists
 of two lines corresponding respectively to
 {\it free-flow} phase (positive slope)
 and {\it jam} phase (negative slope).
It is remarkable that there is a region of density
 where two states (a free-flow state and a jam state) coexist.
The second-order difference allows the model
 to show this property, so-called {\it hysteresis},
 as pointed out in Section 1.

The free-flow line in the fundamental diagram has
 a slope of 1, i.e., all the vehicles are moving
 deterministically (i.e. $v^t_i=1$) without jamming.
This kind of state can be implemented in the uniform state;
 equal spacing of vehicles
 and the initial velocity $v^0_i=1$ for all $i$.
As can be seen from (\ref{SOV}), the intention changes
\begin{equation}
\left\{
\begin{array}{ll}
v^t_i=1-(1-v^0_i)(1-a)^t\quad & (\Delta x^t_i\geq d),\label{v1}\\
v^t_i=v^0_i(1-a)^t & (\Delta x^t_i< d),
\end{array}
\right.
\end{equation}
 and consequently the uniform states constitute a line segment
\begin{equation}
Q=\rho\qquad(0\leq\rho\leq\rho_{\sf h})
\end{equation}
 in the fundamental diagram.
The maximum-flux density $\rho_{\sf h}$, at which the vehicles take
 the minimum value $d$ of equal spacing, is given as follows:
\begin{equation}
\rho_{\sf h}=\frac{1}{1+d}.\label{h}
\end{equation}
In Figure \ref{fd}, we see that
 the formula (\ref{h}) is in the complete agreement with
 the simulated results in the case of $d=2$.

\begin{figure}[t]
\begin{center}
\includegraphics[scale=0.5]{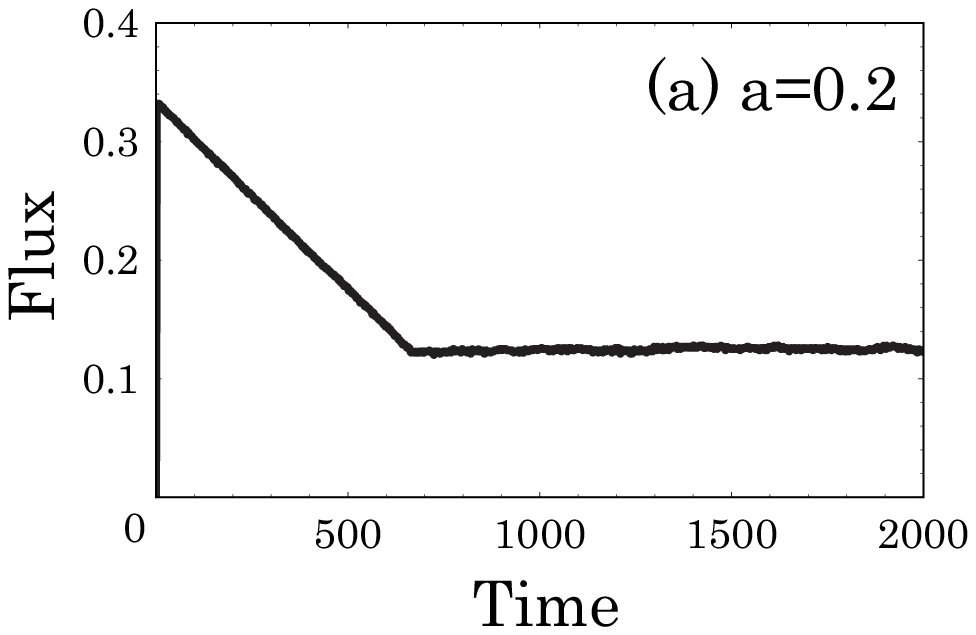}
\hspace{3pt}
\includegraphics[scale=0.5]{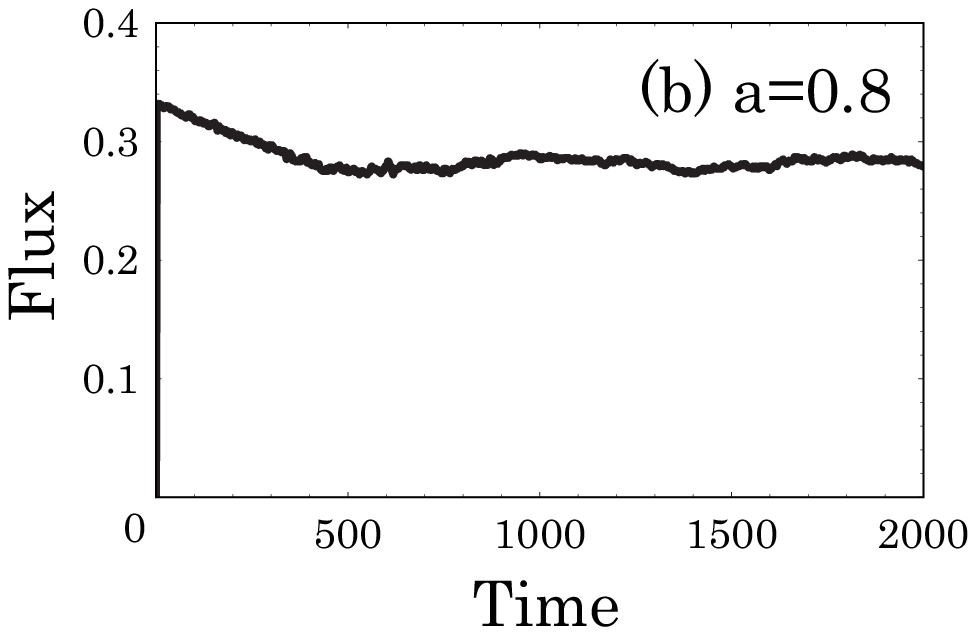}
\caption{
The flux decreases rapidly as an external perturbation is imposed,
 and finally settles into a lower value, i.e., the flux of jam state.
We simulate with the number of sites $L=1000$, the number of vehicles $M=334$, and the sensitivity parameter takes the value of (a) $a=0.2$
 and (b) $a=0.8$.
It imply that, in contrast with our preceding paper,
 the metastable state has no lifetime.
}\label{flux-time}
\end{center}
\end{figure}
The uniform states divide into two classes
 especially under an external perturbation
 which makes a vehicle accidentally slow down.
One is a class of the states recovering their uniform configurations,
 and the other is that of the states never recovering them.
In this paper, we call such a state unstable
 against external perturbations a {\it metastable state}
 in accordance with customary practice.
Figure~\ref{flux-time} shows that, under perturbation,
 the flux of a metastable state is decreasing
 and tends to that of a jam state.
(Note that it also implies that there is not
 such a long-lived metastable state as observed in \cite{KNT}.)
Consequently, in the region of density where two states coexist,
 the uniform states appear as a metastable, higher-flux branch.
\subsection{The critical point of phase transition}\label{criticalpoint}
The flux of traffic flow increases in proportion to
 the density of vehicles while the density is small.
However, as the density becomes bigger,
 close-range interaction between vehicles
 makes a wide, strong correlation over them,
 and consequently gives rise to a jam.
Then, there appears a turning point at which the flux declines
 for the first time.
Around that point (the so-called {\it critical point}),
 the states of traffic flow bifurcates 
 into a stable branch and a metastable branch,
 and moreover phase transition occurs between them.

In order to estimate the critical point,
 we consider that the jam line should be expressed by
\begin{equation}
 Q=\frac{\rho_{\sf c}}{\rho_{\sf max}-\rho_{\sf c}}(\rho_{\sf max}-\rho),\label{jamline}
\end{equation}
 where $\rho_{\sf max}$ and $\rho_{\sf c}$ denote respectively
 the density at which flux vanishes and that of the critical point.
From (\ref{jamline}) and $Q=\rho v$, we have
\begin{equation}
\frac{1-\rho}\rho=\frac{1-\rho_{\sf max}}{\rho_{\sf max}}(1-v)+\frac{1-\rho_{\sf c}}{\rho_{\sf c}}v.\label{headways}
\end{equation}
Equation (\ref{headways}) suggests that
 the spatial pattern divides into two kinds,
 i.e., clustering ($v=0$) and free flow ($v=1$).
Then, since the vehicles move at velocity 1 or 0 in the present model,
 $v$ just indicates the ratio of those
 in free flow, and moreover total average headway
 $\langle\Delta x\rangle=(1-\rho)/\rho$
 is calculated from the average headway of the clustered vehicles
 $\langle\Delta x_J\rangle=(1-\rho_{\sf max})/\rho_{\sf max}$
 and that of the vehicles in free flow
 $\langle\Delta x_F\rangle=(1-\rho_{\sf c})/\rho_{\sf c}$.
These two values reflect a macroscopic property of the SOV model
 with the OV function (\ref{step}), but we should estimate
 them from a microscopic viewpoint.
In what follows, we consider the case of $d=2$,
 following \cite{Sugiyama}.
\begin{figure}[tb]
\begin{center}
\includegraphics[scale=0.7]{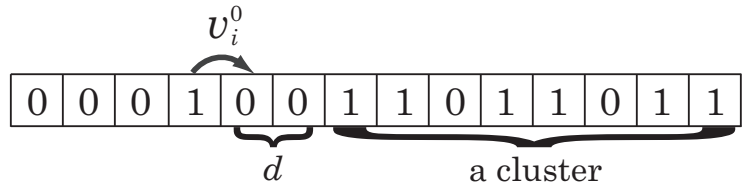}
\end{center}
\caption{
Schematic picture of the case that
 a free vehicle reduces its intention $v^0_i=1$ approaching a cluster,
 and finally comes to be in the cluster.
The vehicle stops with a headway to the cluster ahead,
 which is estimated from the OV function.
}
\label{in}
\end{figure}

First, we think of $\rho_{\sf max}$ as the limit density
 at which all free-flow domains of the roadway close up
 and no vehicle can move then.
Let us consider the situation illustrated in Figure~\ref{in}
 that a free vehicle with its intention 1 is going into a cluster.
Then, taking the time $t=0$ when the headway firstly gets equal to $d$,
 the intention decreases as $v^t_i=(1-a)^t$.
Since it gives the probability of the vehicle moving at $t$,
 the average headway, while in cluster, amounts to
\begin{eqnarray}
\langle\Delta x_J\rangle&=&\prod^\infty_{t=1}(1-v^t_i)\\
&=&\Bigl[\frac{\vartheta_4(0,1-a)^4\vartheta_2(0,1-a)\vartheta_3(0,1-a)}{2(1-a)^{1/4}}\Bigr]^{1/6},
\end{eqnarray}
where $\vartheta_k(u,q)~(k=1,2,3,4)$ are
 the elliptic theta functions.
Consequently, we obtain the density of clustering vehicles
\begin{equation}
\rho_{\sf max}=\frac1{1+\langle\Delta x_J\rangle},
\end{equation}
 as a function of the sensitivity parameter $a$.
Note that $0\leq \langle\Delta x_J\rangle\leq 1$
 since $d=2$, and $\langle\Delta x_J\rangle$ is equivalent to
 the probability of $\Delta x^t_i$ taking the value of 1.

\begin{figure}[t]
\begin{center}
\includegraphics[scale=0.7]{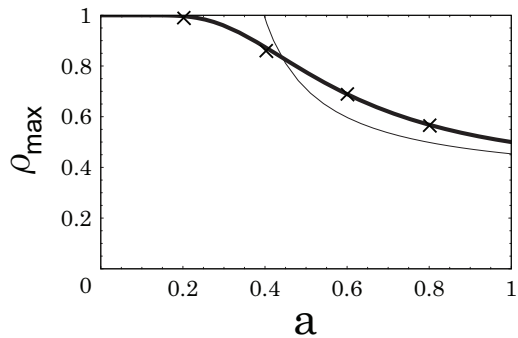}
\end{center}
\caption{
The theoretical curve of the maximum density $\rho_{\sf max}$
 (thick line) at which the flux vanishes
 with the corresponding numerical results (cross).
They have perfect agreement.
We also show a corresponding curve of the original OV model
 (thin line) by use of (\ref{sugi1}).
}
\label{jam}
\end{figure}
Figure~\ref{jam} shows the graph of $\rho_{\sf max}$,
 and it has a perfect agreement with the numerical results
 read off from Figure~\ref{fd}.
In \cite{Sugiyama}, they give the explicit formula of $\Delta x_J$,
 the uniform headway with which vehicles stop in a jam.
It reads
\begin{equation}
\Delta x_J=d-\frac{v_{\sf max}\sigma}{2a},\label{sugi1}
\end{equation}
where $d=2$ and $v_{\sf max}=1$ in the present case,
 and $\sigma\simeq 1.59$.
By use of (\ref{sugi1}), we illustrate the corresponding graph
 in Fig. \ref{jam} as well.
Since the original OV model does not incorporate
 a hard-core exclusion rule,
 the sensitivity parameter $a$ is limited
 in the scope of $\Delta x_J\ge0$
 so as to avoid any collision.
In contrast, the maximum density of the SOV model is retained,
 due to that rule, not to diverge within $0\leq a\leq1$.

\begin{figure}[tb]
\begin{center}
\includegraphics[scale=0.7]{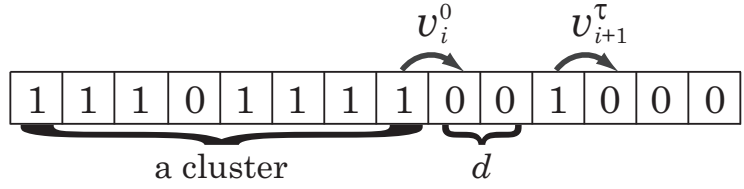}
\end{center}
\caption{
Schematic picture of the situation that
 two adjacent vehicles in the front of a cluster recover their intention
 and get out of the cluster.
We estimate the headway with which the two vehicles finally come to move free.
}
\label{out}
\end{figure}

Next, in order to estimate $\langle\Delta x_F\rangle$
 we consider that two vehicles in the front of a cluster
 are getting out of it as illustrated in Figure~\ref{out}.
Then, as described above, there occur two cases since $d=2$;
 $\Delta x^0_i=1$ with probability $\langle\Delta x_J\rangle$
 and $\Delta x^0_i=0$ with probability $1-\langle\Delta x_J\rangle$,
 where we again take the time $t=0$ when $\Delta x^t_i$ becomes $d$.
Corresponding to $\Delta x^0_i=0$ and $1$,
 we describe $\langle\Delta x_F\rangle$ respectively
 as $\langle\Delta x_F\rangle_0$ and $\langle\Delta x_F\rangle_1$,
 and accordingly our main result is expressed as follows:
\begin{equation}
\rho_{\sf c}=\frac1{1+\langle\Delta x_F\rangle},
\end{equation}
where
\begin{equation}
\langle\Delta x_F\rangle=\langle\Delta x_F\rangle_1\langle\Delta x_J\rangle+\langle\Delta x_F\rangle_0(1-\langle\Delta x_J\rangle).
\end{equation}

Let $\tau$ denote the interval of time for the front vehicle
 to get out of the cluster.
Provided that clusters are large enough to take approximately $v^0_i=0$
 and that the second vehicle leaving the cluster maintains
 a headway of at least $d$, we conclude that
 $v_i^0=0$, $v_{i+1}^0(\tau)=1-(1-a)^\tau$, $v^t_i=1-(1-a)^t$,
 and $v_{i+1}^t(\tau)=1-(1-a)^{\tau+t}$ from (\ref{v1}).
Note that $\tau$ is a stochastic valuable,
 and hence $v^0_{i+1}$ and $v^t_{i+1}$ are dependent on $\tau$.

In the case of $\Delta x^0_i=1$:
 Since $v^t_i$ indicates the probability of moving ahead
 at one site, the probability of $\tau=t$ amounts to
\begin{equation}
P_1(\tau=t)=v^t_i\prod^{t-1}_{s=1}(1-v^s_i)
\end{equation}
and moreover, the distance which the two make in free flow amounts to
\begin{equation}
\sum^\infty_{t=1}\Bigl[v^t_{i+1}(\tau)-v^t_i\Bigr]=\frac{1-a}av^0_{i+1}(\tau).
\end{equation}
Consequently, we have $\langle \Delta x_F\rangle_1$
 in a convenient form for computation:
\begin{eqnarray}
\langle\Delta x_F\rangle_1&=&d+\sum^\infty_{\tau=1}\frac{1-a}av^0_{i+1}(\tau)P_1(\tau)\\
&=&1+\frac{\vartheta_2(0,\sqrt{1-a})}{2(1-a)^{1/8}}.
\end{eqnarray}
\begin{figure}[tb]
\begin{center}
\includegraphics[scale=0.7]{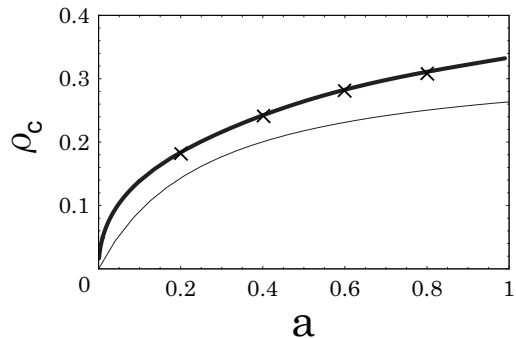}
\end{center}
\caption{
The theoretical curve of the critical density
 $\rho_{\sf c}$ (thick line) at which the flux bifurcates
 into a metastable branch and a stable jam branch (i.e. hysteresis),
 with the corresponding numerical results (cross).
They also have a perfect agreement.
The corresponding curve (thin line) of the original OV model
 is illustrated by use of (\ref{sugi2}).
}

\label{critical}
\end{figure}

In the case of $\Delta x^0_i=0$: The probability of $\tau=t$ amounts to
\begin{equation}
P_0(\tau=t)=v^t_i\sum^{t-1}_{s=1}\Bigl[v^s_i\prod^{t-1}_{r=1,\ne s}(1-v^r_i)\Bigr]
\end{equation}
Consequently, from (\ref{v1}) we have
\begin{equation}
\langle\Delta x_F\rangle_0=d+\sum^\infty_{\tau=1}\Bigl[\frac{1-a}av^0_{i+1}(\tau)P_0(\tau)\Bigr],\\
\end{equation}
and finally $\langle\Delta x_F\rangle$ is formulated
 as a function of the sensitivity parameter $a$.

The formula of $\Delta x_F$ in the corresponding case of the OV model,
 given in \cite{Sugiyama}, reads
\begin{equation}
\Delta x_F=d+\frac{v_{\sf max}\sigma}{2a},\label{sugi2}
\end{equation}
 where $d=2$, $v_{\sf max}=1$, and $\sigma=1.59$.
Figure \ref{critical} shows the critical density $\rho_{\sf c}$
 versus the sensitivity parameter $a$.
It also has a perfect agreement with the numerical results
 read out from Fig. \ref{fd}.
By use of (\ref{sugi2}), we illustrate the corresponding graph
 in Fig. \ref{critical} as well.
We find that the two theoretical curves present
 a qualitative agreement,
 while there are some quantitative differences
 due to the choice of unit car size.
\section{Summary and Conclusion}
The Stochastic Optimal Velocity model was introduced
 in the preceding paper \cite{KNT}
 as a stochastic cellular automaton model
 extending two exactly solvable models
 (the Asymmetric Simple Exclusion Process
 and the Zero Range Process).
Moreover, since it has the same formulation
 as the Optimal Velocity model,
 the SOV model can be regarded as a stochastic extension
 of the OV model.

In the present paper, we take a step function
 as the OV function in order to investigate
 an elementary property of the SOV model.
In previous papers \cite{Sugiyama,Nakanishi},
 the original OV model with this OV function
 are studied in detail and some analytical results are given.
Accordingly, we first make a qualitative comparison between them
 in terms of the motion of each vehicle, and then we see
 a similar hysteresis loop
 in the velocity-headway diagram (phase space)
 as long as the density is low.
That is, as far as each vehicle's motion is concerned,
 the vehicles of the two models show a similar motion
 within a low-density region.

However, as the density of vehicles grows large,
 the structural difference between
 ordinary differential equations and cellular automata,
 as well as the presence or absence of randomness,
 causes crucial differences especially in the fundamental diagram.
As discussed in \cite{Bando}, the metastability observed
 in the OV model is derived from the instability of
 a uniform/homogeneous flow against perturbations.
In general, randomness introduced in traffic models
 tends to eliminate metastable states as well as unstable ones.
Nevertheless, the metastable states appearing
 in the fundamental diagram of the SOV model
 fend off the disturbance of randomness with a trick,
 i.e., the special choice of the OV function and the configuration.
Since that choice allows vehicles to move with probability 1,
 the uniform flow continues to be stable
 while the uniform headway takes the value not less than
 the discontinuous point of the OV function.
That simple consideration leads
 to the formula of the highest-flux density, i.e.,
 the end-point of a metastable branch.
However, all the uniform flows are not entirely stable,
 but higher-flux states of them are metastable,
 i.e., unstable against external perturbations
 as shown in Fig. \ref{flux-time}.
It is one of the most important things
 to evaluate the critical point of density
 where traffic flow becomes unstable
 and clustering of vehicles sets in.

In \cite{Bando2}, they gave the fundamental diagram of the OV model
 with a practical OV function
 and discussed the stability of uniform flow.
Then, they showed the region of density where uniform flows become
 unstable and gave the flux of jammed vehicles formulated analytically  in the fundamental diagram.
In contrast with the original OV model,
 the SOV model incorporates a hard-core exclusion rule,
 and is thus collision-free under any configuration of vehicles.
Moreover, as the intention (or average velocity) is suppressed
 since density is high, the vehicle's motion is controlled
 by randomness as well as the hard-core exclusion rule.
These aspects give the SOV model a fundamental diagram
 which is obvious different from that of the OV model.

We point out some apparent differences from the OV model
 attributed to the effect of hard-core exclusion rule and randomness.
In the fundamental diagram of the SOV model,
 there is a metastable branch with positive slope
 (equal to the maximum velocity 1),
 and the jam line vanishes at a density less than 1.
The simulated results reveal that
 the fundamental diagram of the SOV model with a step OV function
 consists of two line segments with which it has exactly
 the shape of inversed-lambda.
From the diagrams, we conclude that the vehicles move approximately
 with either headway of free-flow or that of jammed vehicles,
 and our issue is thereby reduced
 to estimation of two specific headways.
It can be done entirely by probabilistic calculations,
 and consequently the whole fundamental diagram
 (obviously including the critical point) is successfully formulated
 as a function of the sensitivity parameter.
These analytical results are attributed to the simplicity
 of the SOV model and the choice of OV function.

Further studies on the SOV model under open boundary condition
 and on the multi-velocity version of SOV model
 will be given in subsequent publications \cite{KNT2}.
\ack
The authors appreciate Nimmo J J C for critical reading
 and helpful comments.

This work is supported in part by Grant-in-Aid for Scientific Research
 from the Japan Society for the Promotion of Science (No. 15760047).

\end{document}